# Hyperbolic-enhanced Raman scattering in van der Waals $MoOCl_2$: from Fano resonances to picomolar detection

*Anton Minnekhanov[1†], Gleb Tikhonowski[1†], Georgy Ermolaev[1†], Konstantin V. Kravtsov[1], Gleb Tselikov[1], Adilet Toksumakov[1], Aleksandr Slavich[1], Ivan Kazantsev[1], Ilia Fradkin[1], Andrey Vyshnevyy[1], Ivan Kruglov[1], Ilya P. Radko[1], Zdenek Sofer[2], Aleksey Arsenin[1], Kostya S. Novoselov[3,4,5*], Valentyn Volkov[1*]*

*[1]Emerging Technologies Research Center, XPANCEO, Internet City, Emmay Tower, Dubai, United Arab Emirates*

*[2]Department of Inorganic Chemistry, University of Chemistry and Technology Prague, Technická 5, 166 28 Prague 6, Czech Republic*

*[3]National Graphene Institute (NGI), University of Manchester, Manchester, M13 9PL, U.K.*

*[4]Department of Materials Science and Engineering, National University of Singapore, Singapore, 03-09 EA, Singapore*

*[5]Institute for Functional Intelligent Materials, National University of Singapore, 117544, Singapore, Singapore*

[†]These authors contributed equally to this work

[*]*Correspondence should be addressed to e-mail: kostya@nus.edu.sg and vsv@xpanceo.com*

## Abstract

**Natural van der Waals (vdW) crystals with hyperbolic dispersion challenge artificial metamaterials but remain confined to the mid-infrared. The emergence of $MoOCl_2$, a quasi-one-dimensional metal with in-plane hyperbolicity, overcomes this spectral limit, shifting the focus to the practical visible range. Here, using angle-resolved polarized Raman spectroscopy, we uncover pronounced polarization-dependent Fano lineshapes that reveal coupling between phonons and the anisotropic metallic continuum, together with strong wavelength-dependent reshaping of the Raman polarization patterns governed by the anisotropic optical response of $MoOCl_2$. Harnessing this optical anisotropy, we demonstrate "Hyperbolic-Enhanced Raman" (HypER) scattering, where $MoOCl_2$ provides polarization-tunable analytical enhancement factors exceeding $10^7$ and picomolar-level detection of Rhodamine 6G down to 100 pM, without deliberate surface nanostructuring. These results establish air-stable $MoOCl_2$ as a simple, wafer-compatible platform for visible-range hyperbolic nanophotonics and lithography-free sensing.**

**Keywords:** van der Waals materials, $MoOCl_2$, hyperbolic materials, Raman spectroscopy, SERS.

## Introduction

The manipulation of light at the nanoscale is fundamentally constrained by the diffraction limit,[1] a barrier that hyperbolic materials (anisotropic media in which different principal components of the permittivity tensor have real parts of opposite sign) can overcome by supporting electromagnetic modes with unbounded wavevectors.[2–4] While artificial hyperbolic metamaterials have successfully demonstrated phenomena such as hyperlensing and negative refraction,[2,4] their practical utility is severely hampered by high ohmic losses and complex nanofabrication requirements.[5,6] Consequently, the field has pivoted toward natural van der Waals (vdW) materials, such as hexagonal boron nitride (hBN)[7,8] and α-$MoO_3$,[9,10] which host low-loss hyperbolic phonon polaritons (HPhPs). However, the functionality of these natural crystals is strictly confined to the mid-infrared (mid-IR) spectral window, dictated by the energy scales of their lattice vibrations.[3,11] The search for a natural material capable of supporting in-plane hyperbolic polaritons in the visible range, critical for biosensing and quantum emitters, has remained a long-standing challenge in nanophotonics.[4,5]

This landscape has been transformed by the recent identification of molybdenum oxychloride ($MoOCl_2$), a layered crystal that bridges this frequency gap. Unlike conventional vdW semiconductors, $MoOCl_2$ is a correlated metal featuring quasi-one-dimensional Mo-O chains with orbital-selective Peierls phases.[12–14] This unique electronic structure results in extreme in-plane optical anisotropy: the material exhibits metallic response ($\varepsilon_1 < 0$) along the crystallographic *a*-axis and dielectric response ($\varepsilon_1 > 0$) along the *b*-axis in the visible and near-IR range, enabling it to support in-plane hyperbolic plasmon polaritons (HPPs) across this spectral window.[5,12,13,15,16] Recent real-space nano-imaging has experimentally confirmed that $MoOCl_2$ flakes sustain these visible-range polaritons with remarkably low losses and propagation lengths rivaling those in graphene, all without requiring encapsulation, thereby positioning $MoOCl_2$ as a promising platform for planar nanophotonics.[5,12,16]

Despite the strong interest in its electronic and polaritonic properties, the lattice dynamics of $MoOCl_2$ remain surprisingly unexplored. Typically, Raman spectroscopy of vibrational modes in metals poses a significant challenge.[17] The high surface reflectivity and shallow skin depth of metals drastically limit the interaction volume between light and matter, while the efficient screening of Coulomb interactions by free charge carriers suppresses the modulation of polarizability required for inelastic light scattering.[18] Consequently, Raman spectra of metals are often vanishingly weak or absent. However, in low-dimensional systems such as $MoOCl_2$, strong electronic anisotropy and correlation effects can fundamentally alter this picture, giving rise to unique electron-phonon interaction mechanisms that can be probed by Raman spectroscopy. Furthermore, in low-symmetry 2D materials, Raman spectroscopy reveals intricate light-matter interactions shaped by the crystal's anisotropy.[19] For instance, recent studies on the magnetic semiconductor CrSBr demonstrated that phonon modes can undergo polarization switching: a change in the polarized Raman patterns driven by resonant exciton-phonon coupling.[20] Given the extreme electronic anisotropy of $MoOCl_2$, probing its vibrational response is essential for uncovering similar coupling mechanisms that may govern polariton damping and thermal transport.

In this work, we present an in-depth angle-resolved polarized Raman (ARPR) study of $MoOCl_2$, revealing three remarkable phenomena: (i) pronounced Fano-type lineshape asymmetries along the metallic *a*-axis, that are consistent with strong electron-phonon coupling; (ii) Raman polarization switching with excitation energy; and (iii) anisotropic laser-induced heating. These results demonstrate that the properties that make $MoOCl_2$ hyperbolic also endow it with light-matter coupling. By harnessing this anisotropic response, we introduce "Hyperbolic-Enhanced

Raman" (HypER) scattering, where exfoliated $MoOCl_2$ flakes provide analytical enhancement factors exceeding $10^7$ without any nanofabrication or chemical modification. This establishes air-stable $MoOCl_2$ as a versatile platform for hyperbolic nanophotonics and a robust, scalable substrate for next-generation chemical sensing.

## Results

### Anisotropic Raman response and Fano resonances

Bulk $MoOCl_2$ crystallizes in the monoclinic space group *C2/m* and forms van der Waals (vdW) bonded layers composed of distorted $MoO_6$ octahedra (**Figure 1a**). Within each layer, edge-sharing octahedra form linear Mo-O chains along the crystallographic $a$-axis, which undergo a Peierls-like dimerization and host quasi-one-dimensional metallic states, whereas transport along the orthogonal $b$-axis is semiconducting.[14,15] This extreme in-plane anisotropy is the basis of the predicted hyperbolic optical response of $MoOCl_2$ in the visible range.[5]

To probe how this electronic anisotropy manifests in the lattice dynamics, we performed angle-resolved polarized Raman (ARPR) spectroscopy in a backscattering geometry (**Figure 1b**), where the incident and scattered polarizations ($\hat{e}_i$, $\hat{e}_s$) are controlled by a polarizer and an analyzer. Rotating the laser polarization by an in-plane angle $\theta$ with respect to the laboratory frame maps the full angular dependence of the Raman tensor elements. The false-color intensity map in **Figure 1c** summarizes the ARPR response at 532 nm excitation in a parallel (VV, $\hat{e}_i \parallel \hat{e}_s$) configuration as a function of the polarization angle $\theta$. Spectroscopic ellipsometry was used to initially determine the $a$- and $b$-axes of the flake.[21] Four intense phonon modes at 177, 292, 351, and 431 $cm^{-1}$ dominate the spectra and display a pronounced periodicity, consistent with the Raman-tensors for the $A_g$ irreducible representation of the *C2/m* space group.

A low-energy mode at 123 $cm^{-1}$, however, behaves in a qualitatively different way. As seen in **Figure 1d**, it is clearly resolved in the angle-averaged spectrum, but its intensity is strongly suppressed when the incident beam polarization is aligned with either principal axis ($\vec{E} \parallel a$ (red) and $\vec{E} \parallel b$ (blue)). ARPR maps (**Figure S3**) reveal a characteristic four-lobe pattern of this mode with a 90° periodicity and vanishing intensity along the $a$- and $b$-axes, which matches the expected angular dependence of a $B_g$ mode in a monoclinic crystal with an off-diagonal Raman tensor, leading to $I_{B_g}^{VV} \propto \sin^2 2\theta$.[22] Similar $B_g$ modes have previously been observed in other monoclinic crystals, such as T′-$MoTe_2$.[23]

First-principles calculations reveal the microscopic origin of the Raman-active modes (see **Table 1, Figure 1f,** and **Supplementary Note SN1**). When combined with an analysis of irreducible representations, these calculations show that the experimental peak at 123 $cm^{-1}$ corresponds to the calculated $B_g$ mode at 127 $cm^{-1}$. This peak provides spectroscopic confirmation of the crystallographic frame, since it should be absent in spectra acquired with the incident polarization aligned along either the $a$- or $b$-axis in VV configuration (see **Supplementary Note SN2** for details). However, this is not yet sufficient to unambiguously define the axes in this material using Raman spectroscopy.

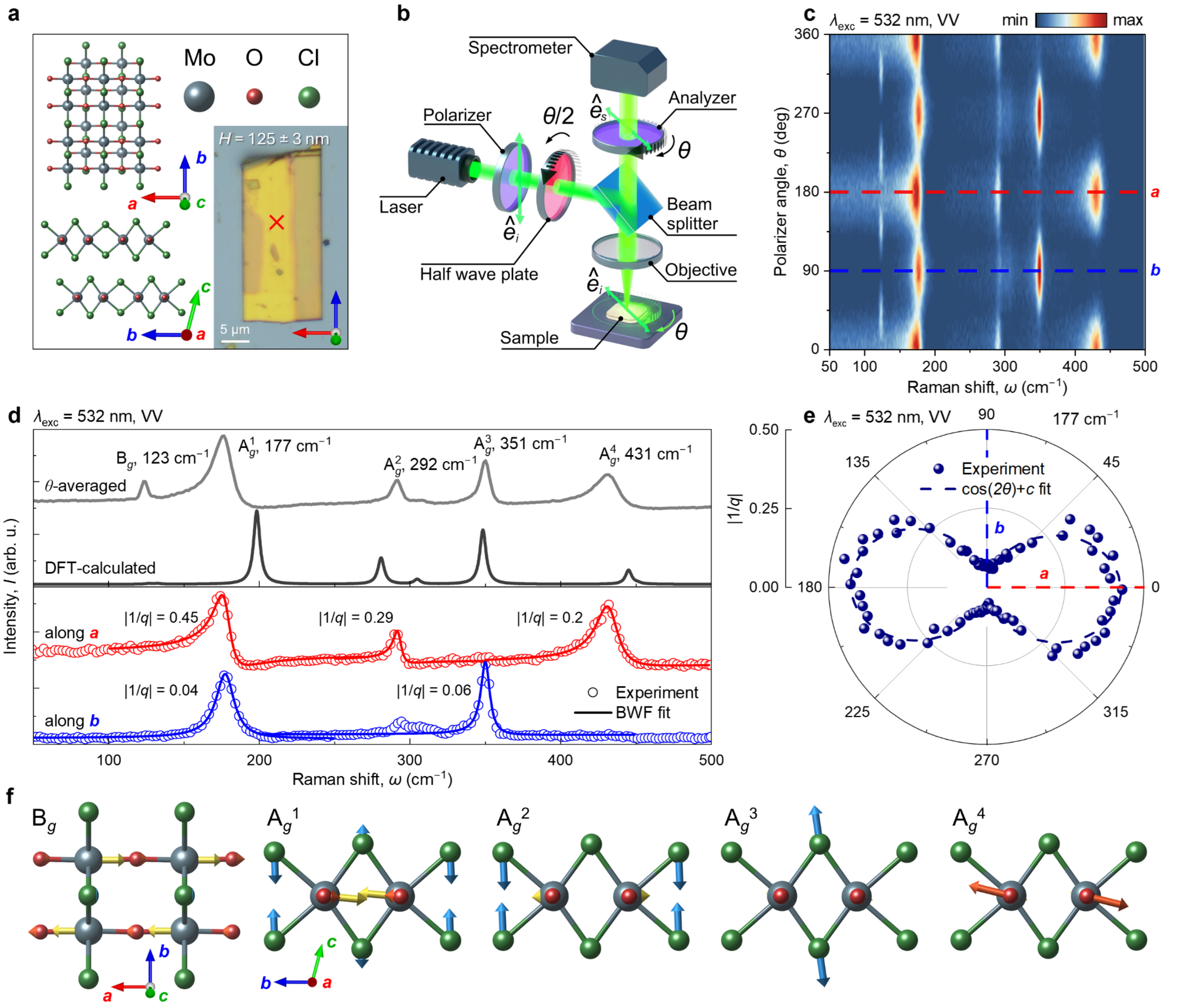

**Figure 1 | Angle-resolved polarized Raman spectroscopy of hyperbolic $MoOCl_2$. a,** Crystal structure of $MoOCl_2$ (space group *C2/m*) showing the Peierls-dimerized chains and crystallographic axes (2 layers are displayed), as well as an optical micrograph of a representative exfoliated flake (the red cross indicates the spot from which the spectra were collected, the thickness of the flake in this area was 125 nm). **b,** Schematic of the backscattering ARPR experiment. The incident beam is linearly polarized, rotated in the sample plane by a half-wave plate, and analyzed in the parallel (VV) configuration. **c,** False-color map of the VV Raman intensity at 532 nm as a function of polarization angle $\theta$ and Raman shift $\omega$. Dashed horizontal lines mark the orientations where the electric field is aligned with the *a*- (red) and *b*- (blue) axes. **d,** $\theta$-averaged experimental and DFT-calculated spectra (top), and spectra for $\vec{E} \parallel a$ and $\vec{E} \parallel b$ (bottom) at 532 nm. For angle-resolved spectra, symbols are experimental data and solid curves are Breit–Wigner–Fano (BWF) fits (Eq. (1)); the extracted asymmetry parameters $|1/q|$ for each mode and orientation are indicated. **e,** Angular dependence of the Fano asymmetry $|1/q|$ for the 177 cm$^{-1}$ mode, showing a two-lobed pattern that maximizes near $\vec{E} \parallel a$ and is strongly suppressed near $\vec{E} \parallel b$; the dashed curve is a cosine fit. **f,** Lattice vibrations, corresponding to the Raman lines in **d**.

Beyond the anisotropy-driven intensity modulation, $MoOCl_2$ exhibits pronounced Raman peak asymmetries that are consistent with direction-selective phonon-continuum interference. When $\vec{E} \parallel b$, all modes in **Figure 1d** (blue spectrum) are accurately described by nearly symmetric profiles, as expected for an insulating dielectric where phonons do not strongly hybridize with an electronic continuum of states. In contrast, spectra recorded with $\vec{E} \parallel a$ (**Figure 1d,** red spectrum) display markedly asymmetric peaks. Such line shapes are characteristic of Fano-type interference between a discrete phonon and a broad electronic continuum.[24,25] We quantify this effect by fitting the peaks with the Breit–Wigner–Fano (BWF) profile:

$$I(\omega) = I_0 \frac{[1 + (\omega - \omega_0)/(q\Gamma)]^2}{1 + [(\omega - \omega_0)/\Gamma]^2}, \tag{1}$$

where $\omega_0$ and $\Gamma$ are the phonon frequency and half-width, and $1/q$ is the dimensionless asymmetry parameter that specifies the magnitude and sign of the Fano asymmetry. **Figure 1d** shows the polarized Raman spectra together with the BWF fits for $\vec{E} \parallel a$ and $\vec{E} \parallel b$, as well as the extracted $| 1/q |$ values for each mode. For the 177 $cm^{-1}$ $A_g^1$ mode, we find $| 1/q | \approx 0.45$ when $\vec{E} \parallel a$, which is reduced by an order of magnitude to $| 1/q | \approx 0.04$ for $\vec{E} \parallel b$. Similar asymmetric features in Raman spectra are consistent with other recent reports for $MoOCl_2$.[26,27]

**Table 1.** Raman modes of $MoOCl_2$.

| $\omega_{exp}$, $cm^{-1}$ | Irreducible representation* | $\omega_{DFT}$, $cm^{-1}$ | Vibration mode |
|---|---|---|---|
| 123 | $B_g$ | 127 | Antiphase gliding (shearing) of the dimerized Mo-O chains along the *a*-axis |
| 177 | $A^1_g$ | 199 | Collective antiphase breathing of all Mo-Cl-Mo-Cl units, including those located between the dimerized Mo-O chains and those positioned between neighboring pairs of such dimerized chains |
| 292 | $A^2_g$ | 281 | Breathing mode involving the Mo-Cl-Mo-Cl units situated between neighboring pairs of dimerized Mo-O chains |
| 351 | $A^3_g$ | 349 | Breathing mode of the Mo-Cl-Mo-Cl units confined between the two Mo-O chains forming a dimerized pair |
| 431 | $A^4_g$ | 445 | Bending (wagging) motion of the O atoms in the O-Mo-O fragments along the *b*-axis |

*Superscripts index the different $A_g$ modes

The angular dependence of $| 1/q |$ for the 177 $cm^{-1}$ $A_g^1$ mode is summarized in **Figure 1e**. The data form a two-lobed polar pattern with 180° periodicity, peaking close to the metallic *a*-axis and reaching a minimum near the semiconducting *b*-axis. This pronounced directionality supports the assignment of the observed Raman line-shape asymmetry to Fano interference involving a strongly anisotropic electronic Raman continuum associated with the quasi-one-dimensional metallic states along the *a*-axis.[14] The maximum in $|1/q|$ therefore provides a Raman marker of the metallic *a*-axis and, together with the angular intensity patterns of the $A_g$ and $B_g$ modes, enables unambiguous assignment of the two in-plane axes under 532 nm excitation.

Fano-distorted phonons have also been reported in the excitonic insulator $Ta_2NiSe_5$ and the quasi-1D superconductor $K_2Cr_3As_3$, where they provide sensitive probes of electron-phonon coupling and collective electronic instabilities.[28,25] In $MoOCl_2$, however, the Fano effect is not tied to a phase transition but is instead intrinsically locked to the one-dimensional metallic channel, providing a Raman signature of its strongly anisotropic electronic response.

## Raman polarization switching

To explore how the angle-resolved polarized Raman response depends on excitation energy, we performed ARPR measurements at three excitation wavelengths: 532, 633, and 785 nm (photon energies of 2.33, 1.96, and 1.58 eV, respectively). The symbols in **Figure 2a** display the angular dependence of the 177 cm$^{-1}$ mode intensity in the VV configuration at these wavelengths. To determine whether this evolution requires a wavelength-dependent Raman tensor, we modeled the air/$MoOCl_2$/Si stack using a birefringence-linear-dichroism/transfer-matrix (BLD/TMM) approach established for Raman scattering in absorbing anisotropic crystals.[22,29,30] The excitation field and the reciprocal collection field at the respective Stokes wavelength were calculated throughout the flake using its measured thickness and independently measured complex refractive indices.[21] These optical inputs were fixed, leaving no adjustable optical parameters. For each mode, one real, symmetric, wavelength-independent effective in-plane Raman tensor was fitted globally to the complete 532-, 633-, and 785-nm VV/VH data set; no tensor element was refitted at an individual wavelength (**Supplementary Note SN2**).

At 532 nm, the polar plot for the 177 cm$^{-1}$ $A_g^1$ mode (**Figure 2a**, also see **Supplementary Note SN2** for other modes and wavelengths) has lobes of comparable intensity along the two axes, forming a 4-lobed "✤"-shaped pattern. Upon lowering the excitation energy to 633 nm and 785 nm, the lobes along the *a*-axis are strongly suppressed and the response becomes *b*-axis-dominated, producing an approximately "8"-shaped pattern. **Figure 2b** summarizes this evolution schematically, while **Figure 2c** shows the corresponding angular cuts. The global model reproduces this wavelength-dependent reshaping without a wavelength-specific Raman-tensor refit.

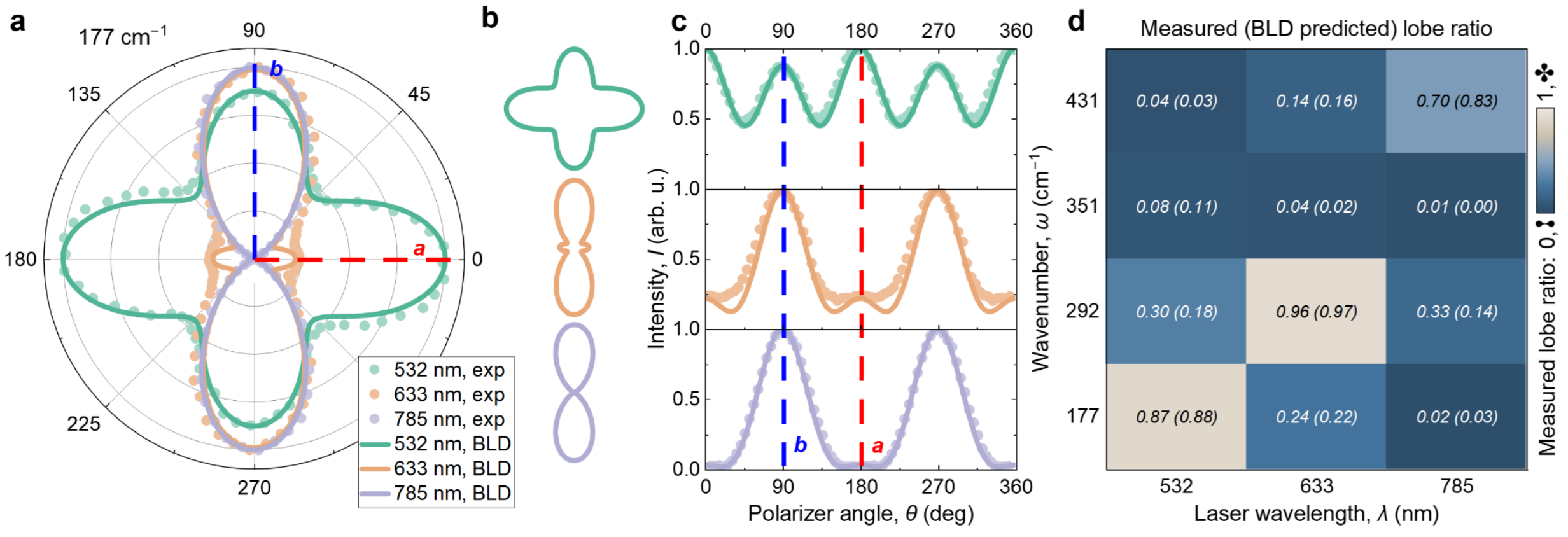

**Figure 2 | Raman polarization switching in $MoOCl_2$. a,** Polar plots of the 177 cm$^{-1}$ $A_g^1$ mode intensity at 532, 633, and 785 nm. Symbols represent experimental data after correction for the

fitted instrumental response (**Supplementary Note SN2**), and solid curves show the BLD/TMM calculation using a single real, symmetric, wavelength-independent in-plane Raman tensor and fixed optical inputs. **b,** Schematic summary of this Raman polarization switching (model curves only). **c,** Corresponding one-dimensional angular cuts. **d,** Heat map of the measured lobe ratio parameter (0 and 1 correspond to 2-lobed and 4-lobed patterns, respectively) for all $A_g$ modes as a function of excitation wavelength (columns) and phonon frequency (rows). Numbers inside the cells list the measured lobe ratio with the model value in parentheses.

Although **Figures 2a–c** display only the 177 $cm^{-1}$ $A_g^1$ mode for clarity, the same global analysis was applied to all $A_g$ modes and to both VV and VH configurations (**Supplementary Note SN2**). **Figure 2d** condenses this angular reshaping through the lobe ratio $I(\theta_{max}+90°) / I(\theta_{max})$, where $I(\theta_{max})$ is the maximum intensity over all angles $\theta$ and $I(\theta_{max}+90°)$ is the intensity measured 90° away from this maximum. Values near 0 (dark blue) correspond to a two-lobe "8"-like pattern dominated by a single axis, whereas values near 1 (light beige) indicate comparable intensities 90° apart, as observed for the four-lobe "✣"-like patterns. Numbers inside the cells in **Figure 2d** list the measured lobe ratios and corresponding model values, showing that the BLD/TMM model captures the observed trends in this parameter. Although the lobe ratio does not encode the full angular shape or its orientation, **Figure S5** shows that the model also reproduces the full angular profiles.

Excitation-energy-dependent reshaping of ARPR patterns has also been observed in several low-symmetry vdW crystals. For example, in CrSBr the principal polarization of an $A_g$ mode rotates by 90° between the crystallographic axes as the excitation energy is varied, a behavior associated with resonant electron-phonon coupling.[20] Excitation-dependent changes in the polarized Raman response have also been reported in $ReS_2$.[31] In black phosphorus, anomalous polarization dependences of the Raman intensities were described by introducing complex (phase-carrying) Raman-tensor components.[32] These observations provide useful context, but they do not imply a common microscopic origin: an intrinsic resonant variation of the Raman tensor must be distinguished from wavelength-dependent propagation and interference in an absorbing anisotropic flake.[22]

This distinction is particularly important for $MoOCl_2$. Spectroscopic ellipsometry and theory show that $MoOCl_2$ has a highly anisotropic electronic and optical response: the *a*-axis exhibits a strong quasi-one-dimensional metallic response, whereas the *b*-axis remains dielectric over much of the visible range. The resulting opposite signs of $\mathrm{Re}(\varepsilon_a)$ and $\mathrm{Re}(\varepsilon_b)$ define the in-plane hyperbolic regime of $MoOCl_2$.[5,12,33] Consequently, the excitation and Stokes fields experience markedly different attenuation, phase accumulation, and penetration depths along the two axes. As shown by the global BLD/TMM analysis above, these measured optical properties are sufficient to reproduce the observed wavelength evolution using one real, symmetric, wavelength-independent effective in-plane Raman tensor per mode (**Supplementary Note SN2**). Related excitation-dependent Raman polarization switching in $MoOCl_2$ was recently reported by Melchioni et al., whose analysis incorporated anisotropic internal fields and finite penetration depth within a TMM treatment.[26] Thus, within the present model, an intrinsic excitation-dependent change of the Raman tensor is not required to explain the lobe switching in $MoOCl_2$, although an additional resonant contribution is not excluded.

Taken together, the ARPR and Fano analyses provide complementary information. The polarization-dependent Fano lineshapes provide evidence for coherent coupling between phonons and a highly anisotropic, quasi-one-dimensional electronic continuum, whereas the wavelength-dependent ARPR reshaping is reproduced by anisotropic optical propagation with a fixed Raman tensor. The latter therefore does not, by itself, establish an intrinsic resonant redistribution of Raman-tensor elements. These complementary signatures provide a spectroscopic fingerprint of the anisotropic electronic response associated with the orbital-selective Peierls state; the same electronic anisotropy underlies the independently established in-plane hyperbolic optical response of $MoOCl_2$.[5,12–14]

## Temperature-dependent anisotropic Raman response

To assess whether the electronic anisotropy of $MoOCl_2$ is accompanied by structural or electronic phase transitions, we first monitored the ARPR response as a function of temperature between 77 and 400 K using 532 nm excitation. **Figure 3a** shows false-color maps of the $\theta$-averaged spectra and spectra acquired with the incident polarization aligned along the crystallographic $a$- and $b$-axes (VV geometry). All Raman-active modes persist over the entire temperature range, with no additional peaks or abrupt changes in their polarization dependence (see **Figure S6** for full temperature-dependent ARPR data for all Raman peaks). These observations provide no evidence for a structural phase transition detectable by Raman spectroscopy and are consistent with the orbital-selective Peierls state previously predicted for $MoOCl_2$ persisting up to at least 400 K.[14]

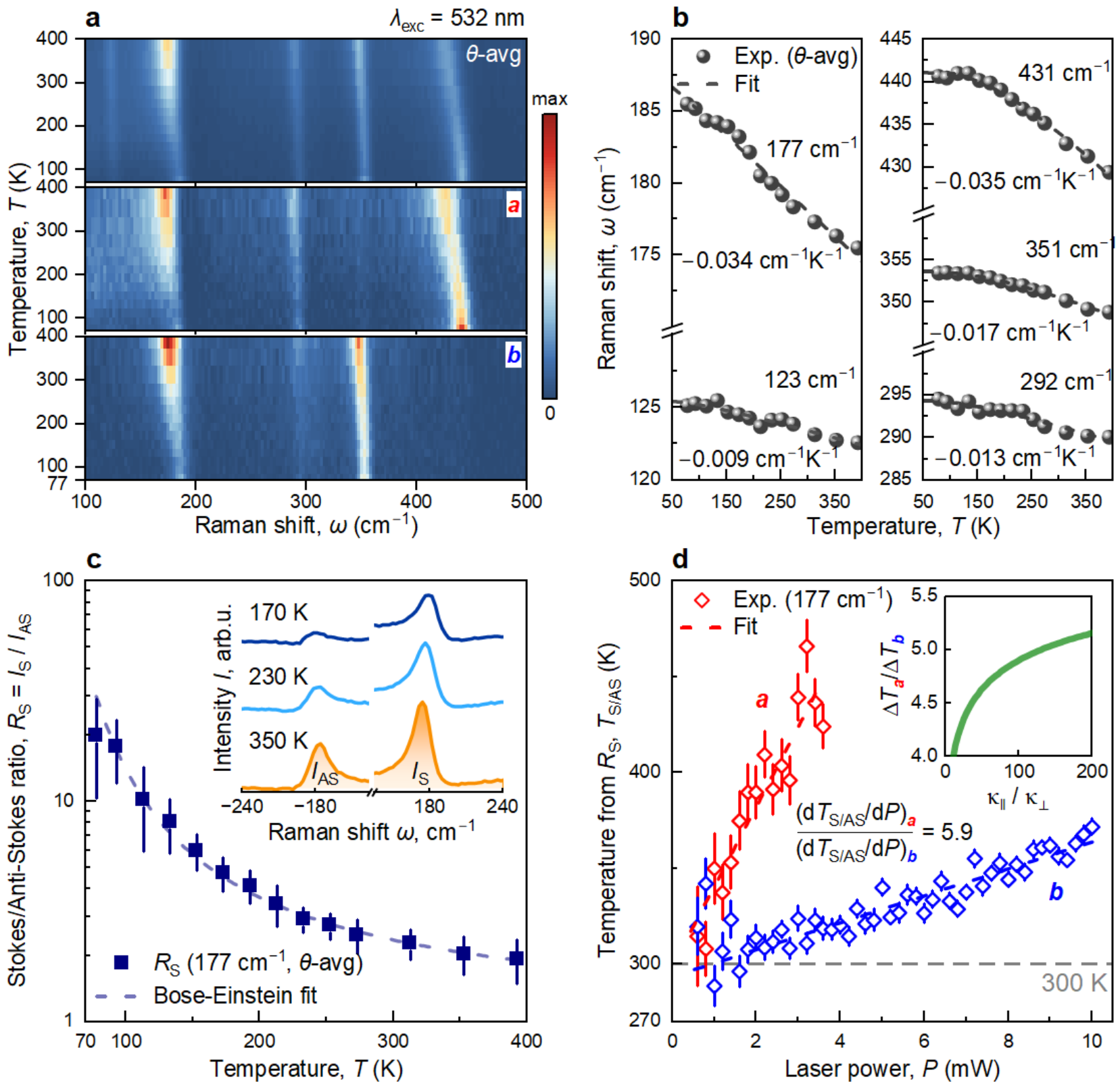

**Figure 3 | Anisotropic temperature-dependent Raman in $MoOCl_2$. a,** Temperature-dependent Raman intensity maps ($\lambda_{exc}$ = 532 nm, VV configuration) for the $\theta$-averaged spectra (top), and for polarizations aligned with the $a$- (middle) and $b$-axes (bottom). **b,** Extracted peak positions of the Raman modes (circles) as a function of temperature ($\theta$-averaged spectra) along with their

corresponding Balkanski fits. **c,** Stokes/anti-Stokes intensity ratio $R_S$ of the 177 cm$^{-1}$ $A_g^1$ mode versus temperature ($\theta$-averaged spectra), together with Bose–Einstein fit (dashed line). Inset, representative Stokes and anti-Stokes spectra. **d,** Effective temperature $T_{S/AS}$ (obtained from $R_S$) versus laser power $P$ for polarizations along $a$ (red) and $b$ (blue). Inset: calculated ratio of the steady-state laser-induced temperature rises, $\Delta T_a/\Delta T_b$, as a function of the thermal conductivity anisotropy $\kappa_{\parallel}/\kappa_{\perp}$ (**Supplementary Note SN4**).

The majority of $A_g$ modes exhibit the expected gradual redshift and broadening with increasing temperature, characteristic of anharmonic phonon-phonon interactions in layered vdW crystals (**Figures 3a,b**). In contrast, the low-frequency ~123 cm$^{-1}$ $B_g$ mode shows only a very weak temperature dependence (**Figure 3b**). Over the measured temperature range, the $\theta$-averaged frequency of the 177 cm$^{-1}$ $A_g^1$ mode softens at an average rate of approximately −0.034 cm$^{-1}$K$^{-1}$, whereas the 123 cm$^{-1}$ $B_g$ mode shifts by only approximately −0.009 cm$^{-1}$K$^{-1}$. Both temperature dependences are adequately described by a standard anharmonic model that includes the leading three-phonon contribution.[34] The nearly fourfold difference in slopes underscores the comparatively weak thermal sensitivity of the $B_g$ mode. Identified as the antiphase gliding of dimerized Mo-O chains along the $a$-axis, this vibration represents a rigid-body shear motion. Unlike the radial deformations of the $A_g$ modes, the restoring force for such inter-chain sliding is governed by shear interactions rather than direct bond stretching. Its distinct eigenvector provides a plausible microscopic basis for the weak thermal renormalization of this mode. Despite their markedly different temperature sensitivities, the frequencies of the remaining Raman modes also evolve smoothly throughout the measured range, without abrupt shifts or other anomalies up to 400 K.

We next use the prominent $A_g^1$ (177 cm$^{-1}$) mode to establish a Raman-based thermometer. **Figure 3c** displays the Stokes-to-anti-Stokes intensity ratio $R_S = I_S / I_{AS}$ obtained from the integrated areas of the BWF fits, as a function of the cryostat temperature $T$. The data follow the expected thermal detailed-balance relation

$$R_S(T) = C\exp(\frac{\Theta}{T}),$$

where $\Theta = \hbar\omega / k_B$ is the characteristic phonon temperature and the temperature-independent prefactor $C$ accounts for the optical-frequency dependence and instrumental response (see **Supplementary Note SN3**). The fitted value $\Theta$ = 268 ± 27 K agrees with the expected value of 253 K for the 177 cm$^{-1}$ phonon, showing that the S/AS ratio follows thermal detailed balance under the low-power calibration conditions. The inset of **Figure 3c** shows representative Stokes and anti-Stokes spectra at selected temperatures, illustrating the systematic evolution of the intensity ratio.

Although the 177 cm$^{-1}$ mode exhibits a pronounced Fano lineshape for $\vec{E} \parallel a$, its use for S/AS thermometry is supported by analogous measurements of additional Raman modes. The S/AS ratios of the 431 cm$^{-1}$ and the 350 cm$^{-1}$ modes are likewise well described by the expected Bose–Einstein temperature dependence for their respective phonon frequencies (**Figure S7**). We therefore use the temperature obtained from the 177 cm$^{-1}$ mode, $T_{S/AS}$, in the laser-heating analysis below.

**Anisotropic laser heating**

Using this calibrated thermometer, we quantify laser-induced heating under 532 nm excitation as a function of incident power $P$ for polarizations along the $a$- and $b$-axes (**Figure 3d**). To remain within the independently established reversible regimes, the analysis was restricted to $P \leq 3.6$ mW for $\vec{E} \parallel a$ and $P \leq 10$ mW for $\vec{E} \parallel b$ (see **Supplementary Note SN4**). For both polarizations, $T_{S/AS}$ increases linearly with $P$ over the measured range, with zero-power intercepts consistent with room temperature: 299 ± 12 K for $\vec{E} \parallel a$ and 294 ± 3 K for $\vec{E} \parallel b$. The extracted heating rates are $dT_{S/AS}/dP$ = 41.7 ± 4.9 K·mW$^{-1}$ for $\vec{E} \parallel a$ and 7.0 ± 0.4 K·mW$^{-1}$ for $\vec{E} \parallel b$, corresponding to a heating rate anisotropy of 5.9 ± 0.8. Independent S/AS temperatures extracted from the other accessible Raman modes yield the same heating rates within uncertainty (**Figure S9** and **Table S4**), demonstrating that this result does not depend on the choice of Raman mode.

The approximately sixfold larger heating rate for $\vec{E} \parallel a$ follows the direction of the strong optical-loss anisotropy of $MoOCl_2$. At 532 nm, spectroscopic ellipsometry gives $k_a/k_b \approx 13$, with the larger extinction coefficient along the metallic $a$-axis.[21] However, the anisotropy of the laser-induced temperature rise cannot be inferred directly from $k_a/k_b$, because both the magnitude and spatial distribution of the absorbed power depend on the internal electromagnetic field, while the resulting $\Delta T$ additionally depends on heat transport. Therefore, we combined transfer-matrix calculations based on the measured optical constants with steady-state heat-transfer modelling of the experimental $MoOCl_2$/Si geometry (**Supplementary Note SN4**).

Although the electromagnetic calculation yields only 41% more total absorbed power for $\vec{E} \parallel a$ than for $\vec{E} \parallel b$, the corresponding heat-source profiles differ markedly. For $\vec{E} \parallel a$, energy deposition is concentrated within the upper ≈ 45 nm of the flake, whereas for $\vec{E} \parallel b$ it extends throughout the flake thickness, resulting in a ten-fold heating power density at the top surface for $\vec{E} \parallel a$. Using representative thermal conductivities for a layered $MoS_2$ vdW crystal, $\kappa_\parallel$ = 100 W·m$^{-1}$·K$^{-1}$ and $\kappa_\perp$ = 2 W·m$^{-1}$·K$^{-1}$, the model gives heating rates $dT_a/dP$ = 33.0 K·mW$^{-1}$ and $dT_b/dP$ = 7.13 K·mW$^{-1}$, corresponding to the ratio of 4.6. These values approximately reproduce the experimental rates for $MoOCl_2$ determined above. The calculated ratio further increases with $\kappa_\parallel/\kappa_\perp$ (**Figure 3d**, inset), showing that the temperature anisotropy reflects both polarization-dependent optical energy release and subsequent heat transport. Thus, the strongly anisotropic dielectric response that underlies the in-plane hyperbolicity of $MoOCl_2$ is also translated into a pronounced polarization-dependent photothermal response under far-field excitation.

Taken together, the temperature-dependent measurements show that the Raman response of $MoOCl_2$ evolves smoothly without evidence of a Raman-detectable phase transition up to 400 K, whereas the power-dependent measurements reveal a pronounced but reversible polarization-controlled photothermal response. The experiments and thermal modelling connect this behavior to the strongly anisotropic dielectric response of $MoOCl_2$. Having established the response of the crystal itself to optical excitation, we next examine whether $MoOCl_2$ can enhance Raman scattering from nearby Rhodamine 6G molecules and consider the possible role of its visible-frequency hyperbolic response.

## Hyperbolic-enhanced Raman scattering

To examine whether the anisotropic metallic and hyperbolic optical environment of $MoOCl_2$ can support enhanced Raman scattering from external analytes, we deposited Rhodamine 6G (R6G) on exfoliated flakes and compared their response to a reference plasmonic substrate. **Figures 4a-c** show an optical micrograph and AFM maps of a representative $MoOCl_2$ flake with a thickness of 218 nm. The probed region exhibits an RMS roughness of only ≈ 1 nm, with no evidence of nanoparticle aggregates, nanogaps, or other surface nanostructures that could act as conventional SERS hot spots.[35] **Figure 4a** defines the area for Raman mapping with the incident polarization aligned along the metallic *a*-axis and the semiconducting *b*-axis, respectively. R6G solutions with concentrations ranging from $10^{-12}$ to $10^{-3}$ M were drop-cast onto the flakes and dried.

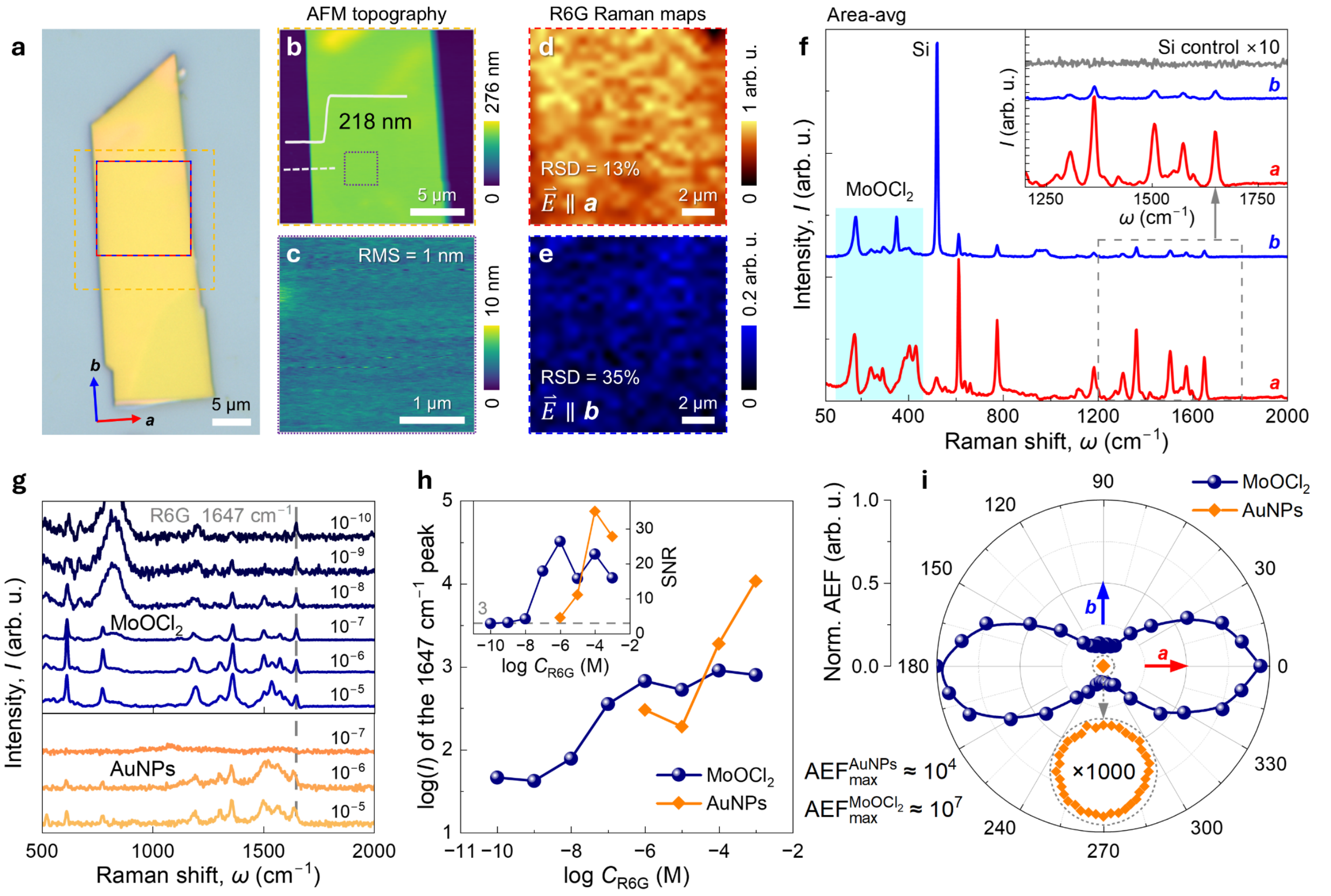

**Figure 4 | Hyperbolic-enhanced Raman (HypER) response of $MoOCl_2$. a,** Optical micrograph of a representative $MoOCl_2$ flake with crystallographic axes indicated and the single area used for Raman mapping (panels **d,e**). The red and blue dashed outlines mark the same region of the flake, mapped consecutively with the incident electric field aligned along the metallic *a*-axis and along the semiconducting *b*-axis. Orange dashed box indicates the area studied using AFM. **b,** AFM topography of the same flake. The dotted purple box marks the region shown in greater detail in **c**. **c,** Zoomed AFM map of the flake showing a root-mean-square roughness of ≈ 1 nm over micrometre-scale areas. **d,e,** Raman maps (24×24 points) of the 1360 $cm^{-1}$ band of Rhodamine 6G (R6G) at a nominal concentration of $10^{-7}$ M, recorded with $\vec{E} \parallel a$ (**d**) and $\vec{E} \parallel b$ (**e**). **f,** Area-averaged Raman spectra from the regions in **d** and **e** for $\vec{E} \parallel a$ (red) and $\vec{E} \parallel b$ (blue) in VV configuration, showing the $MoOCl_2$ phonon modes and the R6G vibrational fingerprints. Inset, zoom of the 1200–1800 $cm^{-1}$ range highlighting the R6G peaks and the Si spectrum averaged from

the control substrate area near the flake, demonstrating no detectable Raman lines (its intensity increased ×10 for clarity). **g,** Representative Raman spectra of R6G (normalized to the 1647 $cm^{-1}$ peak and vertically shifted for clarity) deposited on $MoOCl_2$ (navy) compared with a reference plasmonic substrate formed by Au nanoparticles (AuNPs, orange). The vertical dashed line marks the 1647 $cm^{-1}$ band used for quantitative analysis in **h-i**. **h,** Log-log plot of the 1647 $cm^{-1}$ peak intensity versus R6G concentration for $MoOCl_2$ (navy) and AuNPs (orange). Inset, corresponding signal-to-noise ratio (SNR=$I_{signal}/\sigma_{noise}$). **i,** Polar plot of the normalized analytical enhancement factor (AEF) as a function of incident polarization angle (VV configuration) for $MoOCl_2$ (navy) and AuNPs (orange; scaled by ×1000 for clarity in the inset). All Raman data in **d–i** were acquired using 532 nm excitation.

Even at a nominal concentration of $10^{-7}$ M, Raman maps of the 1360 $cm^{-1}$ R6G marker band reveal a pronounced anisotropy (**Figures 4d,e**). When $\vec{E} \parallel a$, the intensity is high and relatively uniform across the mapped area (RSD = 13%), whereas for $\vec{E} \parallel b$ the signal is reduced severalfold despite identical illumination conditions. This contrast is quantified below using angle-dependent Raman measurements and an anisotropic-Fresnel model. Area-averaged spectra (**Figure 4f**) confirm that the full vibrational fingerprint of R6G is clearly resolved on top of the $MoOCl_2$ phonon lines. Also, in the inset to **Figure 4f**, one can see that no signal is visible against the noise background from the control region of the Si substrate. The nearly featureless AFM topography rule out conventional hot spots associated with nanoparticle aggregates or nanogaps.[36,37] They do not, however, exclude a role of nanoscale roughness in mode out-coupling, considered below.

To assess the polarization anisotropy of the molecular Raman response, we compare $MoOCl_2$ directly with an isotropic reference SERS substrate consisting of colloidal Au nanoparticles (AuNPs) prepared by pulsed-laser ablation in liquid[38] (see **Supplementary Note SN6**) and exhibiting an optical extinction band that overlaps the 532 nm excitation (**Figure S12**). **Figures 4g,h** show representative spectra and the concentration dependence of the 1647 $cm^{-1}$ peak intensity for R6G on $MoOCl_2$ and AuNPs (see also **Figure S13**). The AuNPs exhibit conventional SERS behavior: the peak intensity decreases approximately proportionally with concentration down to $10^{-6}$ M, below which the signal rapidly falls below the noise floor. By contrast, $MoOCl_2$ maintains a clearly resolved Raman fingerprint of R6G down to ≈ $10^{-10}$ M, representing a $10^4$-fold improvement in experimental limit of detection (LOD) compared to the tested AuNPs reference. The inset of **Figure 4h** quantifies this difference by plotting the signal-to-noise ratio (SNR) of the 1647 $cm^{-1}$ peak as a function of R6G concentration. Using SNR ≈ 3 as a conservative detection threshold gives LOD($MoOCl_2$) ≈ $10^{-10}$ M, compared to LOD(AuNPs) ≈ $10^{-6}$ M. The lower LOD on $MoOCl_2$ may also partly reflect the lower R6G fluorescence background relative to the tested AuNP substrate and the resulting reduction in background shot noise (**Figure S11**).

The concentration-normalized analytical enhancement factor (AEF) was calculated using the standard expression:[36]

$$\mathrm{AEF} = \frac{I_{\mathrm{SERS}}/C_{\mathrm{SERS}}}{I_{\mathrm{ref}}/C_{\mathrm{ref}}},$$

where $I_{SERS}$ and $I_{ref}$ are the intensities of the selected R6G Raman band measured on SERS substrate and the reference sample, respectively, while $C_{SERS}$ and $C_{ref}$ are the corresponding concentrations of the R6G solutions used for deposition. The reference was a dried R6G film prepared from a $10^{-2}$ M solution on Si and measured under identical acquisition conditions. Full calculation details and limitations are provided in **Supplementary Note SN5**. For the 532 nm dataset shown in **Figure 4**, this yields AEF(AuNPs) ≈ $10^4$ for the tested AuNPs and AEF($MoOCl_2$) > $10^7$ for $\vec{E} \parallel a$.

The laser-synthesized AuNP substrate used here was not engineered to maximize SERS performance, and optimized plasmonic substrates can achieve enhancement factors comparable to or higher than those measured for $MoOCl_2$.[37] Accordingly, the quantitative AEF and LOD comparison applies specifically to the AuNP reference tested here, whose primary role is to provide an isotropic control for the polarization dependence of the molecular Raman signal. **Figure 4i** summarizes the analytical enhancement factor as a function of the incident polarization angle for $MoOCl_2$ and AuNPs. The AuNPs, as expected, show an almost isotropic response. By contrast, $MoOCl_2$ displays a pronounced "8"-shaped polar pattern, with lobes of maximum enhancement aligned along the crystallographic $a$-axis. The maximum enhancement for $\vec{E} \parallel a$ exceeds that for $\vec{E} \parallel b$ roughly an order of magnitude, providing a simple far-field handle to modulate the analyte signal. This contrast follows directly from the intrinsic anisotropic optics of $MoOCl_2$, which is metallic along the $a$-axis and dielectric along the $b$-axis: an anisotropic-Fresnel model of the air/$MoOCl_2$/Si stack reproduces the observed $a/b$ intensity ratio (≈8–9) through the axis-specific local pump and Stokes fields (see **Supplementary Note SN7**). Such vectorial control and polarization-encoded "on-off" capability are difficult to achieve with conventional plasmonic SERS platforms and are highly attractive for polarization-encoded sensing schemes.[39]

A separate wavelength-dependent series acquired using the protocol described in **Supplementary Note SN8** confirmed that the enhanced Raman response persists at 633 nm, although the analytical performance decreases toward longer wavelengths. Within this series, the AEF at 633 nm was approximately three orders of magnitude lower than at 532 nm for both crystallographic axes, while the experimental LOD for $\vec{E} \parallel a$ increased by approximately three orders of magnitude at 633 nm and nine orders of magnitude at 785 nm relative to 532 nm. Importantly, the polarization dependence was retained at all three excitation wavelengths, with the response consistently stronger for $\vec{E} \parallel a$ (see details in **Supplementary Note SN8**).

To determine whether the observed response is generic to layered vdW surfaces, mechanically exfoliated $MoS_2$ and hBN were examined in a separate comparative series as controls that do not share the in-plane hyperbolic response of $MoOCl_2$ at 532 nm (see **Supplementary Note SN9**). Under matched acquisition conditions, the LOD on $MoOCl_2$ was lower than those on $MoS_2$ and hBN by seven and six orders of magnitude, respectively, for $\vec{E} \parallel a$. Unlike $MoOCl_2$, both controls exhibited isotropic R6G Raman responses. Plasmon-free Raman enhancement of R6G under 532 nm excitation has previously been reported for both materials. In $MoS_2$, it has been associated mainly with photoinduced charge transfer, often assisted by molecular resonance, whereas in hBN it has been attributed to interfacial dipolar interactions and analyte adsorption.[40,41] These controls therefore show that the much stronger, polarization-selective response of $MoOCl_2$ is not a generic consequence of depositing R6G on a layered vdW surface.

## Discussion and Outlook

**Figure 5** contextualizes $MoOCl_2$ among representative Raman-enhancing platforms by comparing reported AEF, LOD, and fabrication requirements. Conventional LSPR-based SERS substrates (Au/Ag nanostructures) routinely reach AEF $\approx 10^4$–$10^8$, but are spectrally narrow and rely on careful nanoengineering. Ag-based substrates can additionally suffer from surface degradation. Semiconductor-, defect- and MOF-assisted platforms provide alternative enhancement mechanisms but often require defect engineering, heterostructure formation, or multistep synthesis, whereas the highest reported AEF values among the systems considered here ($>10^{10}$) are achieved in LSPR-charge transfer (CT) hybrid architectures that still depend on dense noble-metal nanostructuring. $MoOCl_2$ offers a complementary route based on a natural hyperbolic vdW material. Under the present experimental protocol, mechanically exfoliated flakes used without lithography, etching, or nanoparticle deposition provide AEF $\geq 10^7$ and an R6G LOD of $\approx 10^{-10}$ M, together with polarization-tunable enhancement by aligning the incident field along the quasi-one-dimensional metallic *a*-axis. Its distinguishing feature is therefore the combination of strong analytical performance and lithography-free polarization control. Standard mechanical exfoliation and transfer were the only substrate-fabrication steps used here. Translation to scalable formats will require large-area growth or transfer of $MoOCl_2$, but the underlying concept does not intrinsically require nanopatterning.

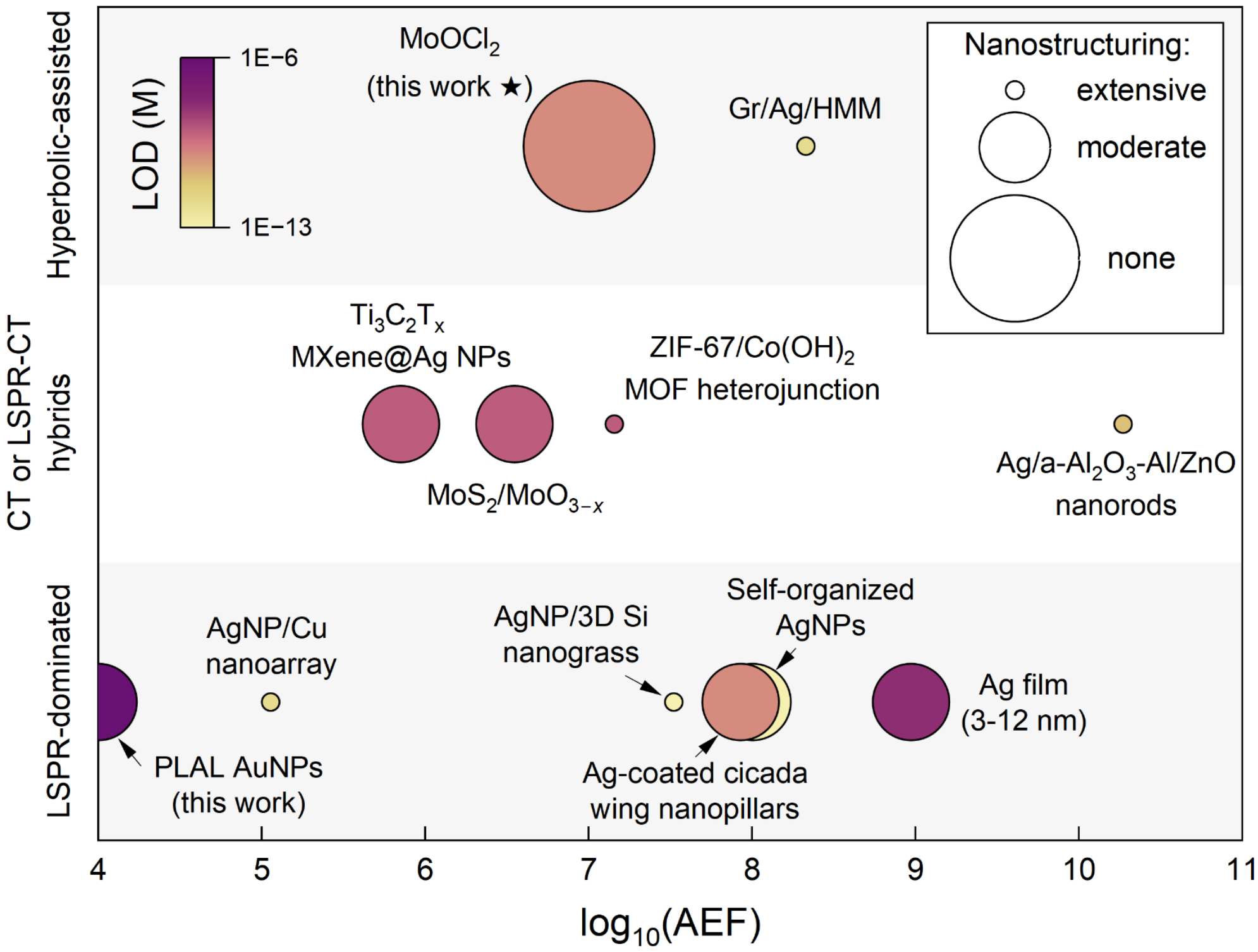

**Figure 5 | Comparison of $MoOCl_2$ with representative Raman-enhancing platforms.** Bubble chart comparing analytical enhancement factors (AEFs) reported for Rhodamine 6G (R6G). The horizontal axis shows $\log_{10}$(AEF), whereas the vertical position is categorical and groups the platforms according to their reported dominant enhancement architecture. The color of each bubble encodes the reported detection limit (LOD) for R6G (color bar). Bubble area is inversely related to the extent of deliberate nanostructuring required: small, medium, and large bubbles correspond to extensive, moderate, and no deliberate nanostructuring, respectively. The comparison includes ultrathin Ag films,[42] self-organized AgNPs,[43] $MoS_2/MoO_{3-x}$ heterostructures,[44] MOF

heterojunctions (ZIF-67/Co(OH)$_2$),[45] Gr/Ag/HMM† architectures,[46] $Ti_3C_2T_x$ MXene@Ag NPs,[47] Ag/a-$Al_2O_3$-Al/ZnO nanorods,[48] Ag-coated cicada-wing nanopillars,[49] AgNPs/3D Si nanograss,[50] and AgNP/Cu nanoarrays,[51] together with $MoOCl_2$ and the PLAL AuNP reference (at 532 nm) from this work. Published values are shown as reported and are not corrected for differences in excitation wavelength, Raman band, acquisition conditions, or analyte adsorption. †Gr/Ag/HMM is included as an artificial-hyperbolic comparator: for this platform, the plotted $x$ value is the reported molecule-number-normalized R6G EF, and the bubble color denotes the lowest demonstrated crystal violet concentration.

We note that the concentration-normalized AEF does not quantify enhancement on a per-molecule basis. Determining a molecule-normalized substrate enhancement factor (SSEF) would require independent quantification of the molecular surface density within the optically sampled region on both the enhancing and reference substrates. Regarding the comparison with AuNP and vdW materials, each control was compared with $MoOCl_2$ under comparable conditions within its respective series. The resulting AEF and LOD differences therefore remain analytically meaningful for the stated protocols (**Supplementary Notes SN6** and **SN9**).

From a mechanistic perspective, our Hyperbolic-Enhanced Raman (HypER) platform differs from conventional LSPR-based SERS platforms. The response correlates with the strongly anisotropic optical and electronic properties of $MoOCl_2$: the same metallic $a$-axis that exhibits Fano-type electron-phonon coupling and anisotropic laser heating also maximizes the Raman enhancement for external molecules. This common directionality links the three observations to the same anisotropic electronic and optical environment but does not by itself establish a shared microscopic mechanism. Hyperbolic media support high-$k$ modes and can substantially modify the projected local density of optical states (LDOS) experienced by a nearby Raman scatterer, thereby affecting both excitation and Stokes emission.[52–55] However, an enhanced total LDOS does not necessarily produce a proportionate increase in the collected far-field Raman signal because part of the emitted energy may be coupled into lossy or non-radiative high-$k$ channels. Nano-optical studies have directly imaged strongly confined hyperbolic plasmon polaritons in $MoOCl_2$ at visible and near-infrared wavelengths and, separately, a long-range anisotropic plasmon mode.[5,27]

A related question is how the enhanced density of high-momentum hyperbolic modes, whose large in-plane wavevectors (and correspondingly short, sub-diffraction effective wavelengths) lie outside the air light cone and cannot be out-coupled through a planar interface, contributes to a far-field-detectable signal. Full-wave calculations (**Supplementary Note SN10**) indicate that the nanoscale surface roughness unavoidable even on nominally smooth exfoliated flakes (RMS $\approx$ 1 nm, **Figure 4c**) can act as a near-to-far-field coupler: a subwavelength perturbation polarized by the strongly enhanced dipole near field re-radiates part of the otherwise trapped energy into the collection cone, and does so anisotropically, favouring the metallic $a$-axis. Because $MoOCl_2$ supports a broadband continuum of such high-$k$ modes rather than a discrete geometric resonance, this out-coupling requires neither engineered nanostructuring nor wavevector-matched features: the wavevector mismatch is bridged locally by roughness-mediated scattering. This is consistent with the featureless AFM topography and with the spatial uniformity of the enhancement, which stands in contrast to the hot-spot-dominated response of rough plasmonic substrates. At the same time, although our full-wave calculations (**Supplementary Note SN10**) already reproduce this out-coupling semi-quantitatively, a complete quantitative decomposition into bulk hyperbolic plasmons, possible surface modes, and residual geometric effects will require further dedicated modelling and nano-optical experiments beyond the scope of the present work.

Taken together, the results establish two distinct levels of interpretation. The polarization contrast of the molecular Raman signal is quantitatively accounted for by the axis-dependent pump and Stokes fields in the $MoOCl_2$/Si stack (**Supplementary Note SN7**). The absolute magnitude of the analytical enhancement, however, is likely multichannel and may include contributions from thin-film interference, LDOS modification and radiative out-coupling, local fields associated with surface perturbations, and possible charge-transfer-mediated chemical enhancement. Substrate-dependent adsorption and retention of R6G can additionally affect the measured AEF, whereas suppression of the fluorescence background can improve the experimental LOD by reducing background-related noise. Separating these contributions and determining a molecule-normalized enhancement factor will require dedicated measurements of molecular surface coverage, fluorescence lifetimes, a dense excitation-energy scan, and local nano-optical fields. The present results therefore establish a reproducible, polarization-tunable analytical enhancement associated with the anisotropic metallic and hyperbolic optical environment of $MoOCl_2$, while leaving its quantitative microscopic decomposition for future work.

From a practical perspective, our results also provide a set of simple and useful guidelines for characterizing $MoOCl_2$ by Raman spectroscopy. First, in the parallel-polarization geometry, the narrow 123 $cm^{-1}$ $B_g$ line provides an internal orientation marker: it is essentially invisible when the incident polarization is aligned with either crystallographic axis, whereas the 351 $cm^{-1}$ $A_g$ mode has maximal intensity when $\vec{E} \parallel b$ and is strongly suppressed for $\vec{E} \parallel a$ under 532, 633, and 785 nm excitation. Moreover, all prominent $A_g$ modes acquire pronounced Fano asymmetry along the metallic $a$-axis. This combination of selection rules and lineshapes enables reliable determination of crystallographic orientation using micro-Raman measurements. The DFT-based assignment of the observed phonon modes to specific vibrational patterns further enables these modes to be used as local probes of strain, defects, and other lattice instabilities that couple selectively to specific atomic displacements. In addition, we show that $MoOCl_2$ remains structurally and vibrationally stable between 77 and 400 K, and identify a practical laser-power window: under our 532 nm, 100× (NA = 0.9) focusing conditions, continuous excitation at powers below ≈ 4 mW ($\vec{E} \parallel a$) and ≈ 21 mW ($\vec{E} \parallel b$) does not produce detectable irreversible changes (**Supplementary Note SN4**). These values should not be interpreted as heating-free operating limits: substantial reversible heating occurs at lower powers, particularly for $\vec{E} \parallel a$, and should be minimized or explicitly accounted for in quantitative Raman measurements. In a separate ambient-stability experiment, no detectable changes in the Raman fingerprint of exfoliated $MoOCl_2$ were observed over 30 days, consistent with previous reports of its stability in air (**Supplementary Note SN11**).[5,27,56]

Thus, the results presented here suggest that hyperbolic vdW metals such as $MoOCl_2$ can serve as simple, robust, and tunable Raman-enhancing platforms that complement traditional noble-metal SERS substrates. The demonstrated stability of $MoOCl_2$ in air, its compatibility with deterministic transfer and stacking, and its intrinsic polarization control together make it attractive for integrated sensing, on-chip spectroscopy, and label-free detection in environments where Ag-based substrates often suffer from surface degradation and poor reproducibility. We anticipate that extending the HypER concept to other natural hyperbolic materials and to more complex vdW heterostructures will open a new direction in natural hyperbolic Raman photonics, where electrically and optically reconfigurable LDOS, rather than lithographically defined hot spots alone, becomes a primary design feature.

## Author Contributions

A.M., G.Tikhonowski, and G.E. contributed equally to this work. A.M., G.E., G.Tselikov, I.R., Z.S., A.A., K.S.N., and V.V. suggested and directed the project. A.M., G.Tikhonowski, G.E., A.T., A.S., I.Kazantsev, and I.R. performed the measurements and analyzed the data. K.V.K., I.Kruglov, I.F., and A.V. provided theoretical support. A.M. and G.E. wrote the original manuscript. All authors reviewed and edited the paper. All authors contributed to the discussions and commented on the paper.

## Competing Interests

The authors declare no competing interests.

## Acknowledgments

The authors thank Dr V. Solovei for assistance in preparing Figure 1b.

## Data Availability

The datasets generated and analyzed during the current study are available from the corresponding author upon reasonable request.

## Methods

### *$MoOCl_2$ Sample Preparation*

Bulk $MoOCl_2$ crystals were purchased from HQ Graphene. Flakes were obtained by micromechanical exfoliation using adhesive tape and subsequently transferred onto Si substrates. To ensure a clean flake-substrate interface, the substrates were pre-cleaned in acetone and isopropanol prior to exfoliation.

### *Raman Spectroscopy of $MoOCl_2$*

Angle-resolved polarized Raman (ARPR) measurements (**Figures 1,2**) were performed using a confocal Raman microscope (alpha300 RA, WITec, Germany). Excitation wavelengths of 532, 633, and 785 nm were used with a 100× objective (Zeiss EC Epiplan-Neofluar, NA = 0.9). Typical acquisition parameters were $t$ = 0.6 s, $a$ = 6, $P$ = 0.7 mW (532 nm), $t$ = 2 s, $a$ = 6, $P$ = 10 mW (633 nm) and $t$ = 2 s, $a$ = 5, $P$ = 10 mW (785 nm), where $t$ is the integration time, $a$ is the number of accumulations and $P$ is the laser power at the sample. The scattered light was dispersed by a 1200 lines/mm grating and detected with a back-illuminated deep-depletion CCD camera cooled to −60 °C, giving a spectral resolution of ~0.5 $cm^{-1}$. Data processing was performed using WITec Project 7 software. Spectra were acquired in parallel (VV) and cross (VH) polarization configurations, with the incident and analyzed polarization aligned parallel and perpendicular to each other, respectively. The incident polarization was rotated in-plane by a half-wave plate, and spectra were collected every 5° from 0° to 360°, yielding 73 angular positions. For each excitation wavelength, the instrumental angular frame was obtained by fitting the dominant $\sin^2[2(\theta - \Delta_\lambda)]$ component of each full-circle $A_g$ VH profile and taking the median $\Delta_\lambda$ over the modes. Corrections of +0.5°, +5.6°, and +7.9° were applied to all modes at 532, 633, and 785 nm, respectively, before the global VV/VH fit and plotting (see **Supplementary Note SN2** for details).

For temperature-dependent ARPR measurements (**Figures 3a–c**), we used 532 nm excitation in VV configuration with $t$ = 1 s, $a$ = 5, and $P$ = 2 mW, a 50× objective (Zeiss LD EC Epiplan-Neofluar Dic 50×/0.55) and a 600 lines/mm grating, resulting in a spectral resolution of ~1 $cm^{-1}$. The sample was mounted in a Linkam T96-S heating/cooling stage for temperature control. Spectra were collected every 10° as the incident polarization was rotated from 0° to 360° (37 angles in total). The temperature was varied from 77 to 400 K in 20 K steps between 77 and 300 K and 50 K steps between 300 and 400 K. A more detailed description of the fitting and data processing procedures can be found in **Supplementary Note SN3**.

For power-dependent Raman measurements (**Figure 3d**), we used 532 nm excitation in VV configuration ($t$ = 1 s, $a$ = 1) with the same 100× objective and 1200 lines/mm grating as above. Measurements were performed with the incident polarization aligned along either the crystallographic $a$- or $b$-axis and the analyzer parallel to the polarizer. The laser power at the sample was varied in 0.2 mW steps from 0.2 to 3.6 mW for $\vec{E} \parallel a$ and from 0.2 to 10 mW for $\vec{E} \parallel b$, thereby restricting the analysis to the independently established reversible-heating ranges. Reversibility was verified by forward-backward power cycling. Further experimental and analysis details are provided in **Supplementary Note SN4**.

***DFT Raman calculations***

First-principles calculations were performed using the Vienna Ab initio Simulation Package (VASP)[57] with the Perdew-Burke-Ernzerhof (PBE)[58] exchange-correlation functional along with Grimme-D3 scheme[59] to account for van der Waals interactions. The core electrons were described using projector-augmented wave (PAW) pseudopotentials, treating Mo 4s, 4p, 4d, and 5s, O 2s and 2p, and Cl 3s and 3p states as valence.[60] A plane-wave cutoff of 550 eV was applied. The internal atomic coordinates were relaxed while keeping the experimental lattice parameters fixed at a=3.755 Å, b=6.524 Å, c=12.721 Å, and $\alpha$=104.86°.[61] For this k-mesh of 6×4×2 subdivisions was employed. Magnetic moments were set on Mo atoms, that turned to 0 μB after relaxation.

Lattice dynamics computations were performed using the Phonopy code,[62] utilizing VASP as an interatomic forces calculator. The supercell (based on the unit cell) of 3×2×1 was used together with a $k$-mesh of 4×4×4.

Raman intensities were evaluated with the Phonopy-Spectroscopy package[63] by using VASP for computing the Raman tensors via the derivative of the frequency-dependent dielectric tensors within the independent-particle approximation for each Raman-active mode. For these calculations, a 9×9×5 k-point mesh for the primitive cell was used together with a Gaussian smearing of 0.1 eV. All peaks were broadened with the Lorentzian function of uniform widths of 5 $cm^{-1}$.

### *Atomic Force Microscopy (AFM)*

AFM imaging was conducted in tapping mode using the WITec alpha300 RA Raman-AFM microscope equipped with a NanoWorld ARROW-FMR probe (75 kHz, 2.8 N/m). The scanning speed, scan size, and resolution were tailored to each study area to ensure optimal visualization quality. AFM image processing was conducted using Gwyddion software.

### *Ellipsometry*

To obtain the optical constants *n* and *k*, the full dielectric tensor of $MoOCl_2$ was experimentally determined using a commercial spectroscopic imaging ellipsometer Accurion nanofilm_ep4 (Park Systems). Measurements were conducted over a broad spectral range from 360 nm to 1700 nm, with a 1 nm step. To maximize sensitivity to in-plane and out-of-plane anisotropies and resolve the parameter correlation between thickness and anisotropic optical constants, a multi-angle, multi-azimuth data acquisition was employed. Ellipsometric parameters ($\Psi$ and $\Delta$) were recorded in rotated compensator mode at multiple angles of incidence ranging from 45° to 57.5° in 2.5° steps at the sample stage rotated azimuthally to align the plane of incidence parallel to the crystallographic *a*- and *b*-axes of $MoOCl_2$ flakes. For modelling $MoOCl_2$ ellipsometry, we used Drude-Tauc-Lorentz/Sellmeier optical model (see Ref.[21] for details).

### *Raman enhancement experiments*

Rhodamine 6G (R6G) powder (Chemsavers) was dissolved in Milli-Q deionized water to prepare a $10^{-2}$ M stock solution and then serially diluted in ethanol down to $10^{-12}$ M. The solutions were drop-cast onto Si substrates bearing $MoOCl_2$ flakes or Au nanoparticles (AuNP) and left to dry in ambient conditions. Control samples were prepared by drop-casting pure Milli-Q water onto identical substrates.

ARPR measurements of R6G on $MoOCl_2$ and AuNPs were performed using the WITec alpha300 RA confocal Raman microscope. For the ARPR datasets shown in **Figure 4i**, we used 532 nm excitation with an integration time $t = 1.5$ s, $a = 7$ accumulations, and a laser power $P = 1.5$ mW at the sample, a 100× objective (Zeiss EC Epiplan-Neofluar, NA = 0.9), and a 600 lines/mm grating. Spectra were acquired in VV configuration (incident and analyzed polarizations parallel). The incident polarization was rotated in the sample plane in steps of 10° from 0° to 360°, yielding 37 angular positions. Single-point spectra (**Figure 4g**) for the concentration series were recorded with 532 nm excitation using $t = 2$ s, $a = 8$, and $P = 2$ mW in VV configuration, with the incident polarization aligned along the crystallographic *a*- and *b*-axes of $MoOCl_2$, and along a fixed laboratory axis for the AuNP substrates. The Raman maps shown in **Figures 4d,e** were acquired over one and the same area of the flake, indicated by the coincident red and blue outlines in **Figure 4a** (12×12 μm), sampled on a 24×24 point grid with 532 nm excitation in VV configuration, using the same objective and grating as above, 0.1 mW laser power, and 1 s integration time. Reference spectra used for the AEF calculation were acquired from a dried R6G film prepared by drop-casting a $10^{-2}$ M R6G solution onto bare Si. Further details of the spectral processing, SNR determination, and AEF calculation are provided in **Supplementary Note SN5**.

### *Au nanoparticles synthesis and characterization*

Au nanoparticles were synthesized by femtosecond laser ablation of bulk Au in 1 mM aqueous NaCl (1030 nm, 400 fs, 40 μJ, 200 kHz; 20 min). The resulting colloid had a lognormal hydrodynamic diameter distribution of 22 ± 16 nm (mode ± FWHM), a plasmon extinction peak at ~530 nm, and a final concentration of 100 µg/mL. Full synthesis and characterization details are provided in **Supplementary Note SN6**.